\documentclass[usenatbib,conference]{basi}
\usepackage[british]{babel}
\usepackage[varg]{txfonts}

\usepackage{rotating}
\usepackage{dcolumn}
\usepackage{graphicx}
\usepackage{latexsym}
\begin{document}
\title[The contagion of star-formation: Its origin]{The contagion of star-formation: Its origin} 
%
\author[Anathpindika, S.]%
       {Anathpindika, S.$^1$\thanks{email: \texttt{sumedh$\_$a@iiap.res.in}}\\
       $^1$Indian Institute of Astrophysics, 2$^{nd}-Block, Koramangala$, Bangalore - 560034, India }

\pubyear{2011}
\volume{00}
\pagerange{\pageref{firstpage}--\pageref{lastpage}}

\date{Received \today}

\maketitle
\label{firstpage}

\begin{abstract}
 Dense pockets of cold, molecular gas precede the formation of stars. During their infancy and later phases of evolution, stars inject considerable energy into the interstellar medium by driving shocks either due to ionising radiation or powerful winds. Interstellar shock-waves sweep up dense shells of gas that usually propagate at supersonic velocities. It is proposed, in this paper, to examine the possibility of dense structure-formation and perhaps, future protostar-formation, in a molecular cloud shocked by such a shell. Here I shall discuss results of a self-gravitating, 3-dimensional, high-resolution simulation using the smoothed particle hydrodynamics.
%
%
\end{abstract}

\section{Introduction}\label{s:intro}
Detailed mapping of nearby star-forming clouds at submm wavelengths within legacy surveys such as the SCUBA2 Gould-belt survey, the \emph{Spitzer}, and more recent Herschel survey of these clouds has revealed their richness in structure. Star-forming clouds occurring at different stages of evolution have various sizes, and shapes. Inter-stellar shocks profoundly affect gas dynamics on a large scale, and are likely to be crucial in producing the observed dense filamentary clouds. I propose to examine this hypothesis numerically.
\section{Description of the problem} 
I shall consider a simple test case where a molecular cloud is shocked by an incident shell moving with a supersonic velocity.
The cartoon in Figure 1 demonstrates the relevant physical details of the problem. The computational domain may be conveniently divided into three regions viz. the intercloud medium (ICM), the standing point at the cloud surface, and the interiors of the shocked cloud marked 1, 2 and 3, respectively. The slab, as shown in this cartoon, moves through the ICM from the left to the right with a precollision velocity, $V_{s}$. It forms a standing shock, labelled 2 in Fig. 1,
after colliding with the cloud surface. The impact of this collision generates reflected waves in region 1, while some straddle the cloud surface and others propagate within the post-collision cloud, the so called transmitted-shock, with a velocity, $V_{c}$, in region 3. The reflected shock-waves have been marked with curly arrows in Figure 1. Pressure in each of the three regions will be denoted by $p$ with a subscript 1, 2 or 3, commensurate with the notation introduced in Fig. 1.

  \begin{figure}
  \begin{minipage}[b]{0.5\linewidth}
  \centering
  \includegraphics[angle=0,width=6cm]{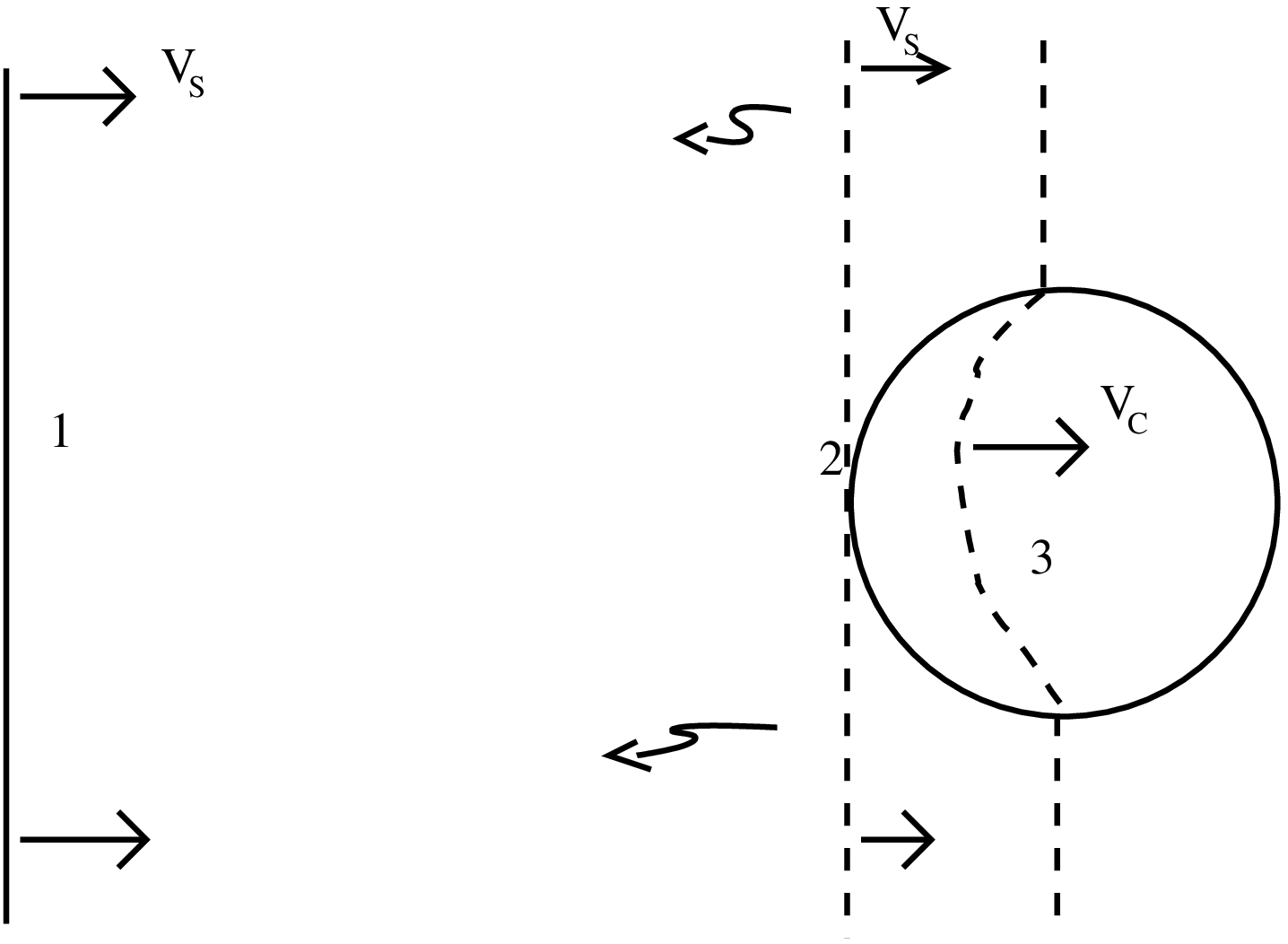}
  \caption{A cartoon showing physical details of the slab-cloud system. See text for description.}
  \end{minipage}
  \hspace{1.cm}
  \begin{minipage}[b]{0.5\linewidth}
  \centering
  \includegraphics[angle=0,width=6cm]{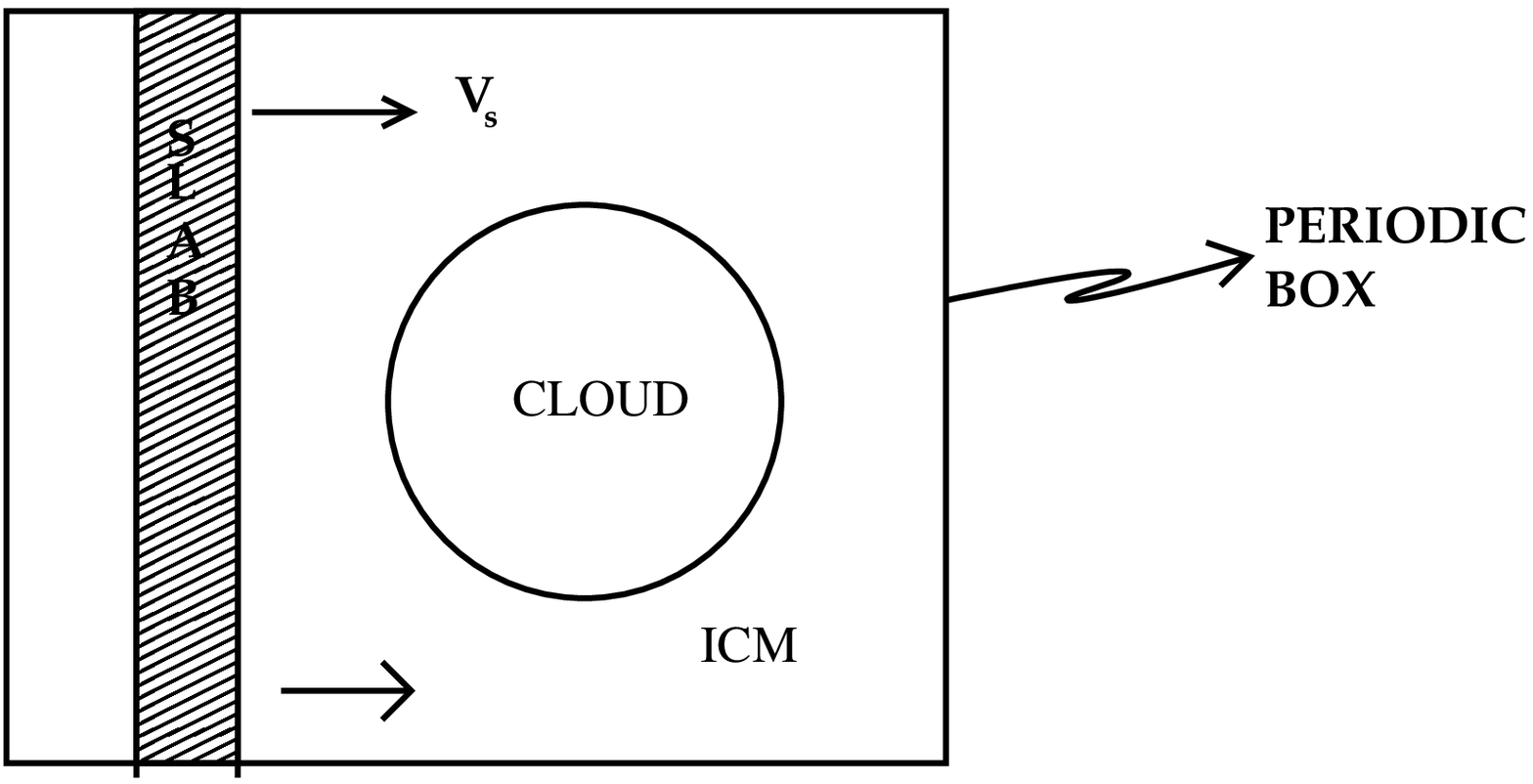}
  \caption{Schematic representation of the computational domain. Shown here is a periodic box enclosing the cloud confined by an intercloud medium, and a slab approaching it with a precollision velocity, $V_{s}$.}
  \end{minipage}
  \end{figure}

Any discussion of this problem would be incomplete without appropriate reference to the shock-dynamics, however, for want of space, I shall restrict myself to only quoting some important results which can be derived using the Rankine-Hugoniot jump conditions (e.g. Courant \& Friedrichs 1956). Following the notation introduced above, the excess pressure due to a highly supersonic shock is
\begin{equation}
\frac{p_{2}-p_{0}}{p_{1}-p_{0}}\sim 2 + \frac{1}{\mu^{2}}, \textrm{where}\ \mu^{2} = \frac{\gamma-1}{\gamma+1}; 
\end{equation}
which for an adiabatic gas constant, $\gamma = 5/3$, produces a pressure-excess of $\sim6$. Pressure within the cloud due to the transmitted shock, $p_{3}$, can be shown to be related to its preshock value, $p_{0}$, as
\begin{equation}
p_{3} = p_{0}\Big[\frac{(1-2M^{2})-1/\gamma}{(1-2M_{+}^{2}) - 1/\gamma}\Big],
\end{equation}
where $M_{+}$ is the Mach number for the incident shock. Observe that, relative to the incident Mach number, there is a fall in pressure at the surface of incidence, the standing point. Finally, the density within the shocked cloud compares with that in the external medium as,
\begin{equation}
\frac{\rho_{3}}{\rho_{0}} = \frac{(\gamma-1)-p_{3}/p_{0}(3\gamma-1)}{p_{3}/p_{0}(\gamma-1) - (3\gamma-1)}.
\end{equation}
For a finite value of this ratio, the velocity of transmitted shock, $V_{c}$, within the cloud can be calculated using the equation of continuity. The resulting effect will be discussed in Section 5 below.
\section{Numerical scheme}
We have used the Lagrangian, particle-based scheme, Smoothed particle hydrodynamics (Monaghan 1992), hereafter referred to as SPH. An SPH particle, in the strictest sense, is not a point particle but has a finite spatial extent defined by a quantity called the smoothing length, $h$. Each SPH particle, characterised by state properties of density, velocity and temperature, interacts with other particles through numerical viscosity. The density of a particle, $\rho_{i}$, is calculated by summing over contributions from nearest neighbours, $N_{neibs}$, of that particle within a radius 2$h$,
\begin{equation}
\rho_{i}\equiv\rho(\mathbf{r}_{i}) = \sum_{j=1}^{Neibs} m_{j}W(\mathbf{r}_{i}-\mathbf{r}_{j},h),
\end{equation}
where $\textbf{r}$ is the position vector of the particle with identifier, $i$. The density of the fluid, $\rho_{s}$, is then simply
\begin{equation}
\rho_{s} = \int W(\mathbf{r} - \mathbf{r'},h)\rho(\mathbf{r'})d\mathbf{r'}
\end{equation}
 The search for the nearest neighbours, and the calculation of net force on an individual particle is done using a tree-algorithm (Barnes \& Hut 1986). Local inhomogeneities are smoothed out using a kernel that has been normalised to unity. Gravity for vanishingly small inter-particle separations is smoothed out using a modified kernel that introduces an infinitesimally small repulsive force between closely spaced particles. The Thomas-Couchman kernel, a special type of the M-4 cubic spline, ensures a finite repulsive force between closely interacting particle-pairs (Thomas \& Couchman 1992). The numerical algorithm employed here, Seren, is a well tested code (Hubber \emph{et al.} 2011), and includes all the features described above. 

The thermodynamic details of the problem are modelled using a simple barotropic equation of state (EOS), defined by Equation (6) below, that mimics the post-collision temperature-jump, $T_{ps}$, and cools off to the precollision temperature, $T_{iso}$, at relatively higher densities.
\begin{equation}
\frac{P}{\rho} = (k_B/\bar{m})\times
\left\{ \begin{array}{ll}
\Big(\frac{T_{ps}}{\textrm{K}}\Big)\ ; \rho\le10^{-23} \textrm{g cm}^{-3}\\
\Big(\frac{T_{cld}}{\textrm{K}}\Big)\ ; 10^{-23}\textrm{g cm}^{-3} <\rho
\le 10^{-22}\textrm{g cm}^{-3}\\
\Big(\frac{\gamma T_{cld}}{\textrm{K}}\Big)\Big(\frac{\rho}{10^{-22}\textrm{g cm}^{-3}}\Big)^{\gamma-1}; 10^{-22}\textrm{g cm}^{-3}\\
< \rho\le 5\times 10^{-22}\textrm{g cm}^{-3}\\
\Big(\frac{T_{cld}}{\textrm{K}}\Big)\ ; 5\times 10^{-22}\textrm{g cm}^{-3}<\rho \\
\le 10^{-18}\textrm{g cm}^{-3}\\
\Big(\frac{T_{iso}}{\textrm{K}}\Big)\Big[1 + \gamma\Big(\frac{\rho}{10^{-15}\textrm{g cm}^{-3}}\Big)^{\gamma-1}\Big]\ ; \\
\rho > 10^{-18}\textrm{g cm}^{-3};
\end{array} \right.
\end{equation}
where the adiabatic gas constant, $\gamma=5/3$, $k_{B}$ and $\bar{m}$ are respectively the Boltzmann constant, and mean molecular mass. 
\section{Initial conditions}
 We propose to test the above theory using the scheme outlined in Figure 2. Ordinary SPH particles, representing gas within the cloud and the slab, interact with each other via gravitational and hydrodynamic forces. The ICM confining the cloud, however, is represented by special particles that exert only thermal pressure on other particles. The entire assembly, including the slab, is enclosed in a self-wrapping periodic box, where the periodicity is limited to merely ghosting particles; in other words, a particle leaving through one face of the box enters from the opposite face. 

Elsewhere in the literature, we have discussed an ensemble of cases for different choices of resolution (Anathpindika \& Bhatt 2011). In this paper though, the emphasis being on the formation of dense structure within the shocked cloud, I will only discuss the simulation with highest resolution.  The precollision cloud was modelled as a sphere having uniform density, and the probability distribution function (PDF) of particles in the virgin cloud peaking at the predicted density, $\rho_{true}$, is shown in Figure 3, which not only demonstrates the stability of the precollison cloud, but also the absence  of any spurious dense pockets.\\
  \begin{figure}
  \begin{minipage}[b]{0.5\linewidth}
  \centering
  \includegraphics[angle=270,width=5cm]{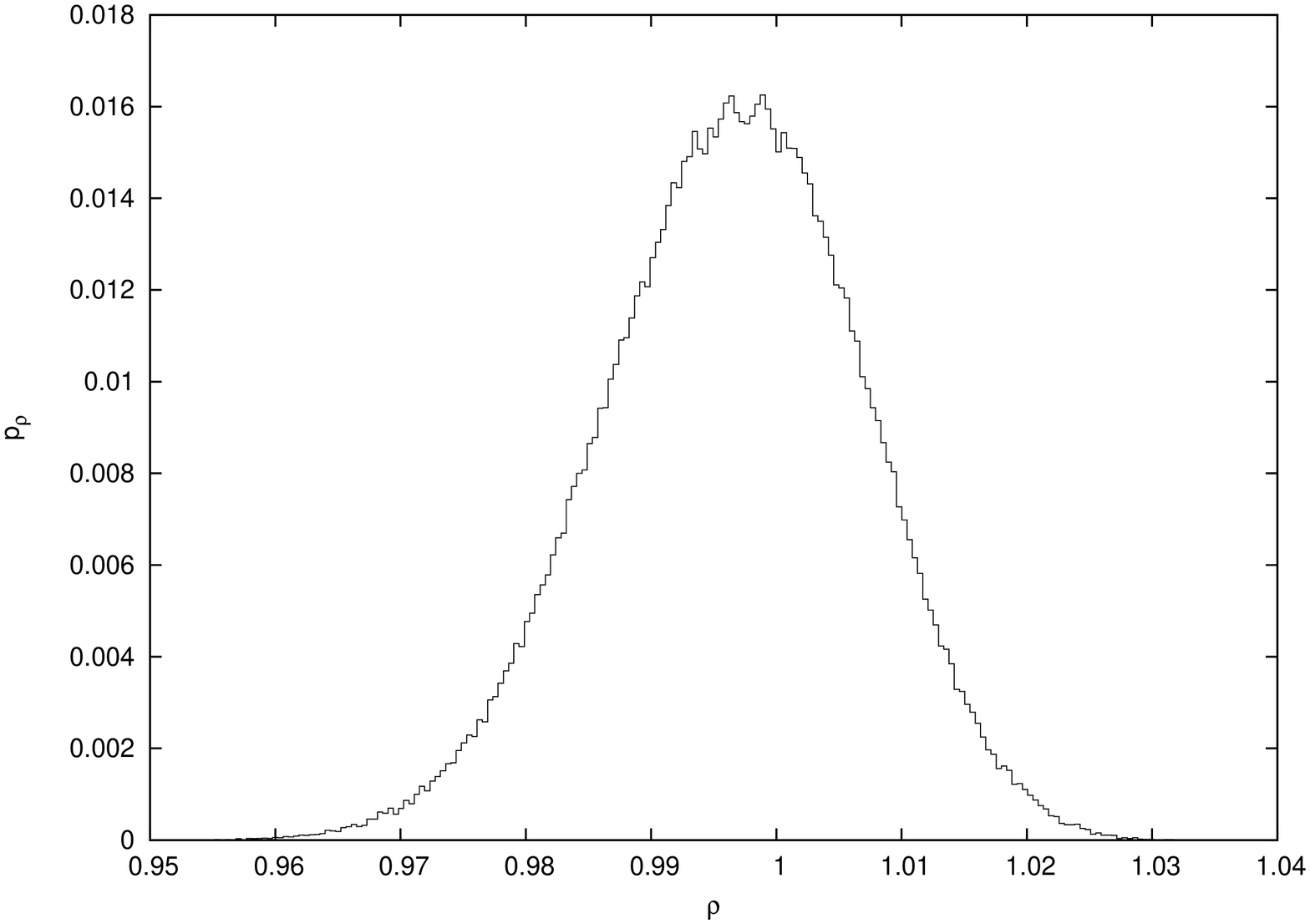}
  \caption{A histogram of the gas density within the precollision cloud. The relatively small spread around the expected density, $\rho_{true}$, indicates the stability of this cloud.}
  \end{minipage}
  \hspace{1.cm}
  \begin{minipage}[b]{0.5\linewidth}
  \centering
  \includegraphics[angle=0,width=5cm]{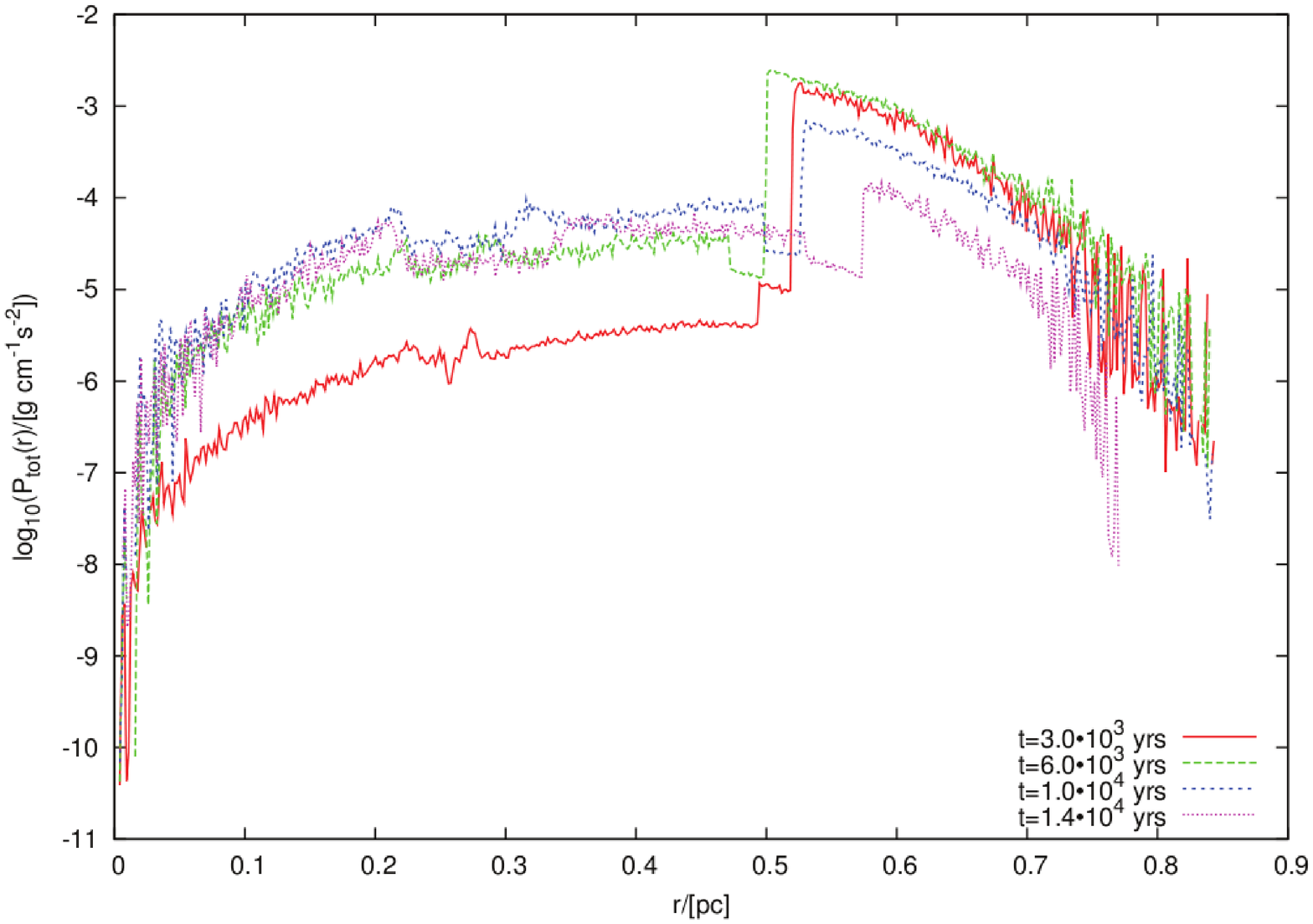}
  \caption{A plot showing the radial variation of pressure within the shocked cloud at different epochs.}
  \end{minipage}
  \end{figure}
\textbf{Important parameters } Mass of the cloud, $M_{cld}$ = 500 M$_{\odot}$; radius, $R_{cld}$= 0.5 pc; uniform temperature, $T_{iso}$ = 15 K.
 The cloud, ICM, and the slab were represented by SPH particles having three different choices of mass, $m$. If $i$, $j$, and $k$ are the respective identifiers of particles in each of the three regions then, $(m_{i}, m_{j}, m_{k})/[\textrm{M}_{\odot}]\equiv(2.2\times 10^{-3},2.53\times 10^{-6}, 1.26\times 10^{-5})$, and the number of particles in each of these regions is respectively (2.4, 2.3, 7.3)$\times 10^{5}$, so that total number of particles, $N_{tot} = 1.2\times 10^{6}$. \\
\textbf{Resolution} The smallest resolvable spatial scale in an SPH calculation, as noted above, is the average smoothing length, $h_{avg}$, so that the volume of a typical SPH particle, $V_{SPH}=\frac{32\pi h_{avg}^{3}}{3}$; and $N_{cld}(32\pi h_{avg}^{3})/3 = 4\pi R_{cld}^{3}N_{neibs}/3$. Thus, 
\begin{equation}
h_{avg}\sim \frac{1}{2}\Big(\frac{N_{neibs}}{N_{cld}}\Big)^{1/3}R_{cld},
\end{equation}
is the typical smoothing length of a particle in the test cloud. In the present case, $h_{avg}\sim 1.5\times 10^{-2}$ pc, and $2\mathcal{R}\equiv\lambda_{J}/h_{avg}\sim 13$, where $\lambda_{J} = (\pi a_{0}^{2}/G\rho)^{1/2}$, is the length of the fastest growing unstable mode in a gas body, the Jeans length. The quantity $\mathcal{R}$ defines the number of SPH particles available to resolve the unstable mode, which in this case is $\sim 6$, and therefore satisfies the Truelove criterion of spatially resolving the instability (Truelove \emph{et al.} 1998).

 \begin{figure}
  \centering
  \includegraphics[angle=270,width=12cm]{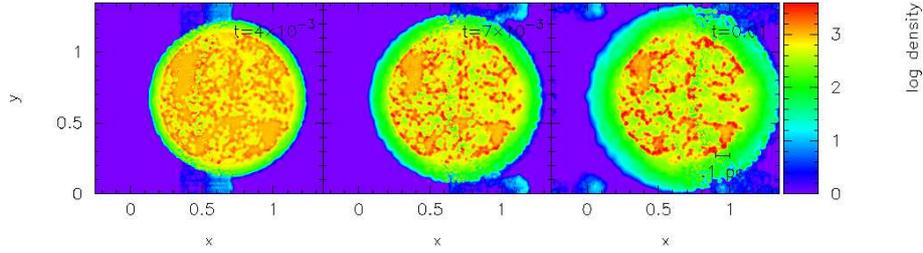}
  \caption{A rendered density plot showing a time sequence of the mid-plane of the shocked cloud, as seen along a direction orthogonal to the plane of the figure. Turbulence induced by the transmitted shock generates clumps and filaments visible in these plots.}
  \end{figure}
\section{Result} 
The post-collision reflected shock-wave as well as the transmitted shock within the cloud is evident from the plot in Figure 4, that shows the radial variation of pressure at different epochs. The incidence of the front surface of the slab on the cloud causes a jump in pressure at its surface, defined by Equation (1), followed shortly by another, relatively smaller, jump due to the rear surface of the slab shocking the cloud (red-curve). The green curve shows the pressure distribution at a slightly later epoch when the pressure-jump moved leftward relative to its position in the red curve, suggesting an outward propagation of a wave, i.e. the reflected shock moving in the ICM. It can be identified as the flared region around the shocked cloud shown in the rendered density plot of Figure 5.

The weaker transmitted shock that is most important in the evolution of the post-collision cloud, initially causes slight contraction of the cloud, evident from the collection of denser gas in a ring on its periphery. Relative to the incident shock, gas within the cloud is at a lower pressure manifested by a trough in the pressure distribution at the surface of incidence, defined by Equation (2). Inward propagation of this shock generates a turbulent velocity field within the cloud that soon produces fractal structure, in other words the appearance of relatively dense pockets of gas interspersed with rarefied regions, the so called holes. Structure within the shocked cloud grows on a timescale comparable to the crossing time, $t_{c}\sim 2R_{cld}/V_{s}$, of the precollision slab which is much shorter than the growth time, $t_{g}$, of the fastest growing unstable mode of length, $\lambda\sim (\pi v_{eff}^{2}/G\rho_{cld})^{1/2}$, where $v_{eff} = a_{0}+v_{c}$; $v_{c}\sim V_{s}/\mathcal{M_{+}}\sim 0.4$ km/s, and $a_{0}(T=20 K)\sim$ 0.27 km/s, so that $\lambda\sim$ 0.1 pc, which implies, $t_{g}\sim$ 0.14 Myr.

The simulation was terminated when the slab, having traversed the width of the cloud, reached the opposite face of the periodic box. Consequently, further evolution of clumps and other contiguous filaments could not be investigated in this work. We have observed that the incident slab, after colliding with the cloud, suffers severe ablation by the time it reaches the other end of the cloud. The transmitted shock, relative to the incident slab, propagates at a much lower velocity within the post-collision cloud, evident from the rendered density plots in Figure (5). However, as argued in Section 2 above, Equations (8) below show that the transmitted shock in a cold, dense gas could still be supersonic, and therefore, will likely have a significant effect on the internal structure of the cloud. The pressure behind the slab changes on a very short timescale, the dynamical timescale, $t_{d}$, of the slab which can be shown to be
\begin{equation}
t_{d}\sim\frac{R_{cld}}{V_{s2}}\frac{(\gamma + 1)^{2}}{\mathcal{M}(\gamma - 1)},
\end{equation}
where
\begin{displaymath}
V_{s2}\sim \Big(\frac{\rho_{0}}{\rho_{3}}\Big)V_{s1}\sim V_{c},
\end{displaymath}
and $\mathcal{M}\equiv \mathcal{V}_{c}/a_{0}$, is the Mach number for the transmitted shock; see paper I for the derivation. In the present case, $t_{d}\sim 10^{-4}$ Myrs $\ll t_{c}$, implying, the dynamical properties behind the shock must indeed change rapidly. In this light the problem under consideration here, in effect, reduces to one of a weak-shock impinging on a cloud or conversely, a shock interacting with a large cloud. \\
  \begin{figure}
  \begin{minipage}[b]{0.5\linewidth}
  \centering
  \includegraphics[angle=270,width=6cm]{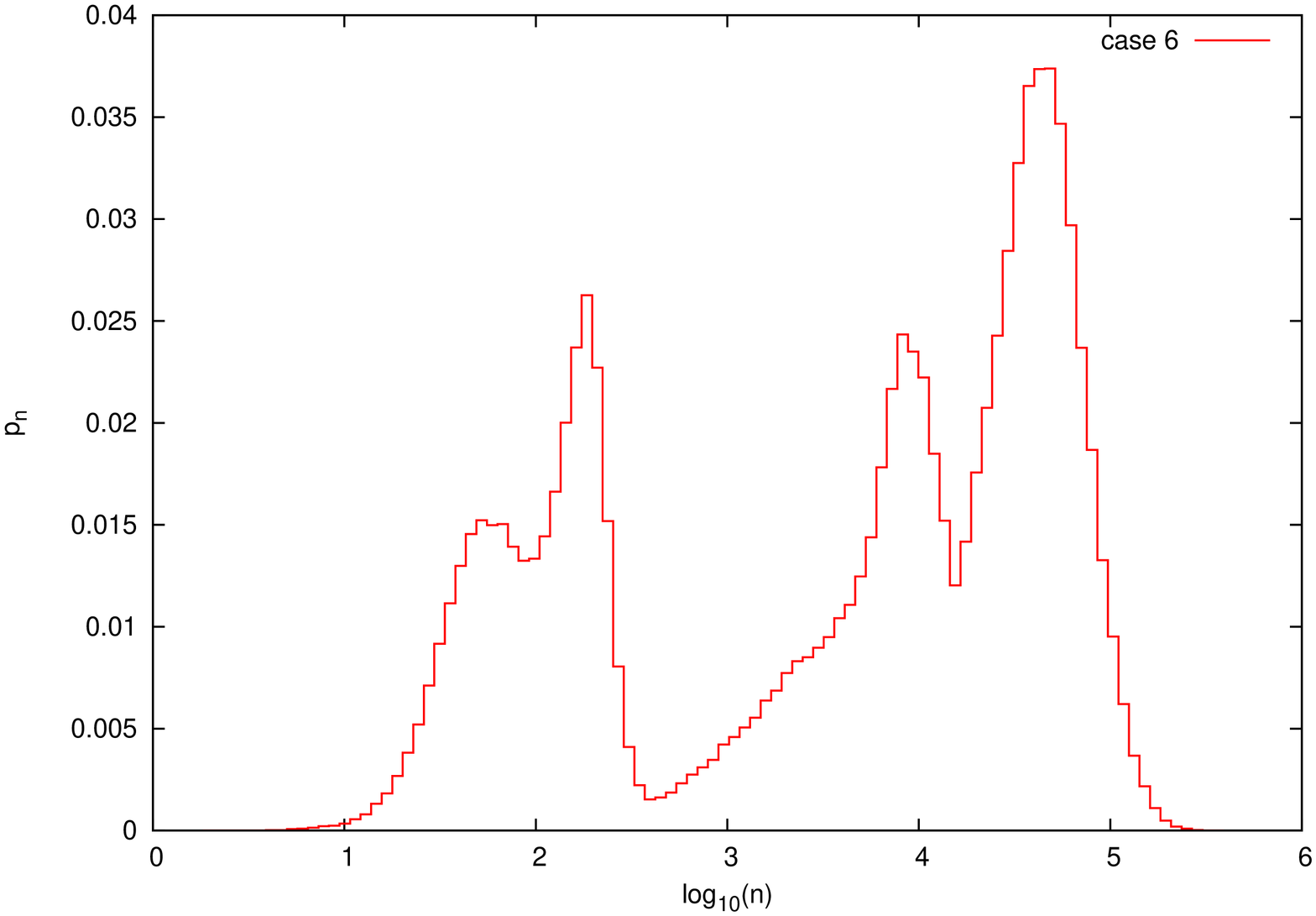}
  \caption{The density PDF for gas in the shocked cloud after the slab has traversed the cloud. Bimodal nature of the density distribution is evident from this plot.}
  \end{minipage}
  \hspace{1.cm}
  \begin{minipage}[b]{0.5\linewidth}
  \centering
  \includegraphics[angle=270,width=6cm]{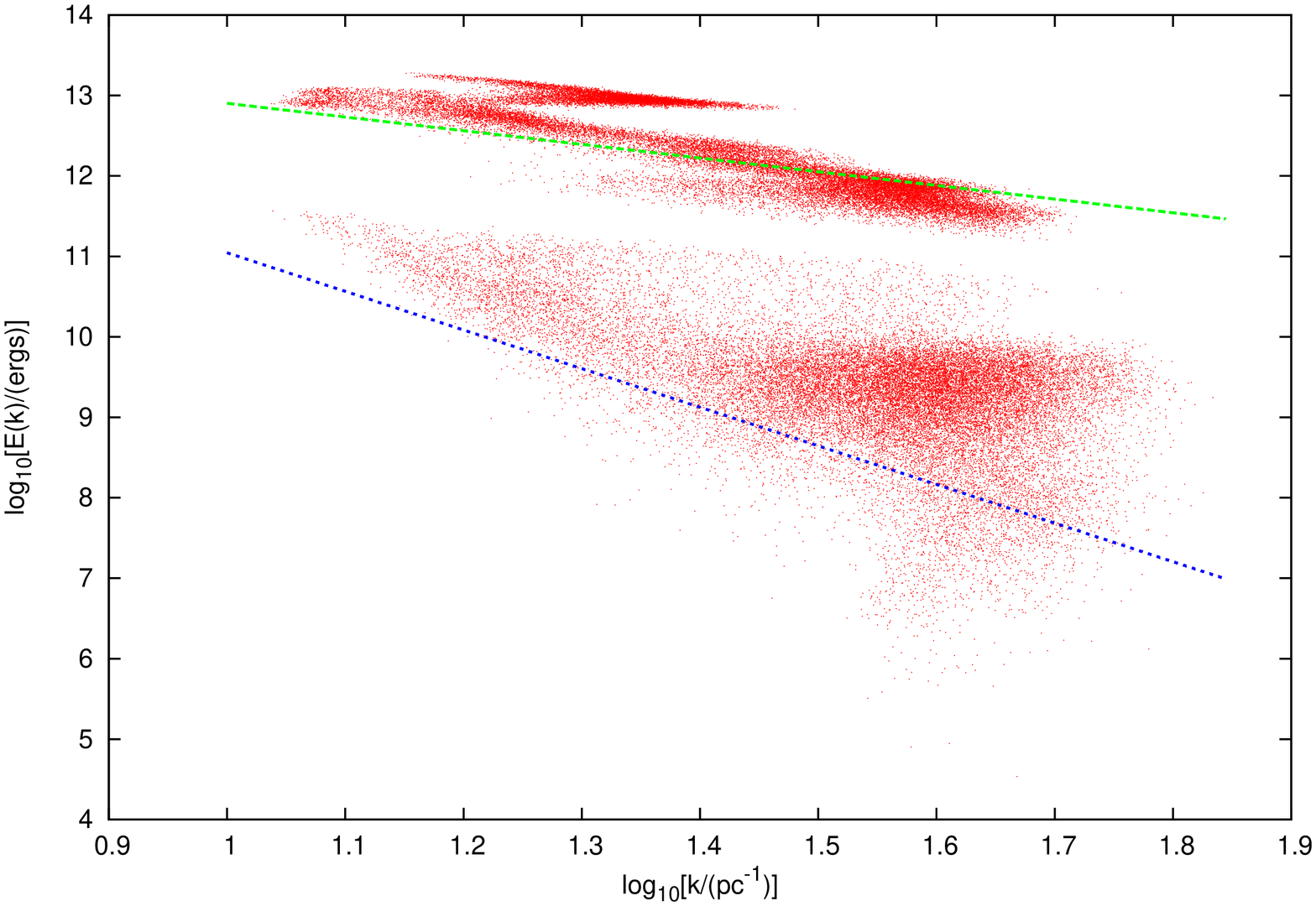}
  \caption{This the corresponding energy spectrum of gas within the shocked cloud. Like the PDF, this spectrum too shows a distinct segregation. The dotted lines show respective power-law fits; see text for description.}
  \end{minipage}
  \end{figure} 
\textbf{The probability distribution function (PDF)}
The CMF has often been suggested as the likely progenitor of the stellar IMF, and the likely relation between the two distributions has been examined by several authors (e.g. Anathpindika 2011, Hatchell \& Fuller (2008), Nutter \& Ward-Thompson 2007, Motte \emph{et al.} 1998). This proposition makes an investigation in to the origin of the CMF even more pertinent. Numerical simulations like those discussed by V{\' a}zquez-Semadeni (1994), and Padoan \& Nordlund (2002), among several other authors, and the PDF plotted in Figure 6 demonstrate the crucial role of interstellar shocks in generating a lognormal density PDF; although we note that the plot shown here is only semi-lognormal. Interestingly, a lognormal fit has also been attempted for the distribution of masses of dense ($\gtrsim 10^{4}$ cm$^{-3}$) cores in nearby star-forming clouds (e.g. Enoch \emph{et al.} 2008). While the apparent similarity of the dense-phase PDF derived here, with the CMF may perhaps be only a fortuitous coincidence, the likelihood of a causal relationship must be further investigated. Also, in the dense phase, as can be seen in Figure 6, this PDF peaks at $\sim 10^{5}$ cm$^{-3}$ which at 15 K, the precollision temperature within the cloud,  corresponds to a characteristic mass, $M_{s}\sim 3$ M$_{\odot}$; incidentally, using a gas-to-star conversion efficiency of 10\% this characteristic mass corresponds to 0.3 M$_{\odot}$, the mass at which the stellar IMF develops a knee before turning over in to the brown-dwarf regime. This test simulation therefore seeks to re-emphasise the importance of interstellar shocks in forming putative star-forming pockets within molecular clouds, and may hold the key to explain the CMF. \\
\textbf{Power-spectrum}
 The energy density, $E(k)$, of a turbulent velocity field in the wavevector ($k$) space, the Fourier domain, is related to its Cartesian counterpart through a simple integral,
\begin{equation}
\int_{0}^{\infty}E(k)dk = \frac{1}{2}\sum_{i} v_{i}^{2}.
\end{equation}
The integral on the left-hand side extends over all possible wave-vectors, while the summation on the left-hand side runs over all SPH particles. The resulting power-spectrum for gas within the shocked cloud has been shown in Figure (7), which like the PDF, is also segregated. Although the spectrum for both, the dense as well as the rarefied phase of the gas, is a power-law it is, however, considerably steeper ($\propto k^{-4.5}$) for the former, and Kolmogorov-like ($\propto k^{-1.7}$) for the latter. The Kolmogorov-spectrum, as is well known, applies to an inviscid, incompressible fluid, though the gas here is viscous, but the rarefied phase appears to obey this approximation; it obviously breaks down for the dense phase.
\section{Conclusions}
\begin{enumerate}  
\item I have argued in favour of my hypothesis that interstellar shocks could possibly lead to formation of dense clumps, and contiguous filaments in molecular clouds.
\item  It has been demonstrated that propagation of a shock renders the density field unstable, and generates structure in it on a rather short timescale. Stars likely to form in these pockets, via energetic feedback, may further inject energy within the gas and quench star-formation in one region, only to trigger it elsewhere.
\end{enumerate}
\textbf{Ancillary remarks } 
 This test case only examined the balance between self-gravity and thermal pressure as the magnetic field was not included, however, magnetohydrodynamic simulations by for e.g. Padoan \& Nordlund (2011) suggest a considerable modulation of the rate at which protostellar objects form in turbulent gas. Other issues deserving a brief explanation include the effect of numerical resolution, and hydrodynamic instabilities on the shocked cloud. The latter is essentially related to the numerical resolution, and more critically, to the influence of SPH viscosity on dynamically unstable fluid layers. First, a relatively poor resolution tends to suppress dynamical fragmentation. Second, the shearing interaction between the slab and the surface of the cloud is also likely to be unstable to hydrodynamic instabilities such as the Kelvin-Helmholtz instability. The demand on resolving this thin layer, however, is conservatively large and not fulfilled here. It does not though, compromise the arguments presented in favour of the hypothesis examined here.


\label{lastpage}
%
\end{document}